\documentclass{aa} % for a referee version
\usepackage{graphicx,longtable}
\begin{document}
\title{Planetary companions around the K giant stars 11~UMi and HD~32518}

% \subtitle{}

   \author{
        M.P. D\"{o}llinger\inst{1}
          \and
        A.P. Hatzes\inst{2}
          \and
        L. Pasquini\inst{1}
          \and
   E.W. Guenther\inst{2}
          \and
        M.~Hartmann\inst{2}
             }

   \offprints{
    Michaela P. D\"{o}llinger \email {mdoellin@eso.org.de}\\$*$~
    Based on observations obtained at the
    2m Alfred Jensch telescope at the Th\"uringer
    Landessternwarte Tautenburg.
  \\}

     \institute{
European Southern Observatory, Karl-Schwarzschild-Strasse 2, {\bf D-}85748
Garching bei M\"{u}nchen, Germany
         \and
        Th\"uringer Landessternwarte Tautenburg,
                Sternwarte 5, D-07778 Tautenburg, Germany}

   \date{Received; accepted}

\abstract
  % context heading (optional)
  % {} leave it empty if necessary
{11~UMi and HD~32518 belong to a sample of 62~K~giant~stars that has been
observed since February~2004 using the 2m~Alfred~Jensch~telescope of 
the Th{\"u}ringer Landessternwarte ($TLS$) to measure precise 
radial velocities (RVs).}
  % aims heading (mandatory)
{The aim of this survey is to investigate the dependence of planet formation
on the mass of the host star by searching for planetary companions around
intermediate-mass giants.}
  % methods heading (mandatory)
{An iodine absorption cell was used to obtain accurate RVs for this study.}
  % results heading (mandatory)
{Our measurements reveal that the RVs of
11~UMi show a periodic variation of 516.22~days with a semiamplitude of
$K$~$=$~189.70~m\,s$^{-1}$. An orbital solution yields a mass function of
$f(m)$~$=$~(3.608~$\pm$~0.441)~$\times$~10$^{-7}$~solar~masses~($M_{\odot}$)
and an eccentricity of $e$~$=$~0.083~$\pm$~0.03.
The RV curve of HD~32518 shows sinusoidal variations with a
period of 157.54~days and a semiamplitude of $K$~$=$~115.83 m\,s$^{-1}$.
An orbital solution yields an eccentricity, $e$~=~0.008~$\pm$~0.03 and 
a mass function, $f(m)$~=~(2.199~$\pm$~0.235)~$\times$~10$^{-8}$~M$_{\odot}$.
The $HIPPARCOS$ photometry as well as our H$\alpha$ core flux measurements 
reveal no variability with the RV period. Thus, Keplerian motion is the most 
likely explanation for the observed RV variations for both giant stars.
}
 % conclusions heading (optional), leave it empty if necessary
{An exoplanet with a ``minimum mass'' of
$m \sin i$~=~10.50~$\pm$~2.47~Jupiter~masses~(M$_{\mathrm{Jup}}$) orbits the 
K~giant 11~UMi. The K1~III giant HD~32518 hosts a planetary companion with a 
``minimum mass'' of $m \sin i$~=~3.04~$\pm$~0.68~M$_{\mathrm{Jup}}$
in a nearly circular orbit. These are the 4th and 5th planets published from 
this $TLS$ survey. 
}
\titlerunning{Planetary companions around 11~UMi and HD~32518}
\maketitle

%
%________________________________________________________________
\section{Introduction}
To date around 350 exoplanets have been detected mostly via the RV 
technique. 
However, these surveys are giving us a very biased view of the 
process of planet formation because less than 10 $\%$ of the planets orbit 
host stars with masses $>$~1.25~M$_{\odot}$. In the search for planets over a 
wider range of stellar masses an increasing number of RV searches are looking
for planets around low- and intermediate-mass stars 
that have evolved off the main sequence (MS) and up the giant branch.
Hatzes $\&$ Cochran (1993) found first indications of substellar companions
around giants. The first extrasolar planet around the K~giant HD~137759 
($\iota$~Dra) was discovered by Frink et al. (2002). Other exoplanets around 
HD~13189 and $\beta$~Gem were detected by Hatzes et al. (2005, 2006). The last 
planet was independently announced by Reffert at al. (2006).\\
Starting in 1998 Setiawan et al. (2003a) began to search for planets around 
83~giants with $FEROS$. This programme detected two giant exoplanets around 
HD~47536 (Setiawan et al. 2003b), one around HD~11977 (Setiawan et al. 2005),
and more recently one around HD~110014 (de Medeiros et al. 2009). We started a 
similar survey in February~2004 (D\"ollinger 2008) monitoring a sample of 
62~K~giant~stars using higher RV accuracy at $TLS$. During this survey planets 
around the K~giants 4~UMa (D\"ollinger et al. 2007), 42~Dra and HD~139357 
(D\"ollinger 2009a) were discovered. In this paper we present 
precise stellar RVs for two other programme stars, 11~UMi and HD~32518, which 
most likely host extrasolar planets in almost circular orbits.\\ 
Moreover several other surveys are actively searching for planets around giant 
stars. Sato started in 2001 a precise Doppler survey of about 300~G--K 
giants (Sato et al. 2005) using a 1.88~m telescope at Okayama Astrophysical
Observatory. From this survey planetary companions around HD~104985 
(Sato et al. 2003), the Hyades giant $\epsilon$~Tau (Sato et al. 2007),
18~Del, $\xi$~Aql, and HD~81688 (Sato et al. 2008) were detected. Furthermore,
this survey discovered planetary companions around 14~And and 81~Cet
(Sato et al. 2008). In the same paper the detection of exoplanets
orbiting the subgiants 6~Lyn and HD~167042 were reported. Niedzielski et al. 
(2007) discovered an exoplanet to the K0~giant HD~17092 using observations 
taken with the Hobby-Eberly Telescope ($HET$) between 2004~January and 
2007~March. Johnson et al. (2007) published exoplanets around the three
intermediate-mass subgiants HD~192699, HD~210702, and HD~175541. Planetary
companions around two other subgiants HD~167042 and HD~142091 were discovered
monitoring a sample of 159 evolved stars at Lick and Keck Observatories for
the past 3.5~years by Johnson et al. (2008).\\ 
Until now five main, yet preliminary results have emerged from the 
$TLS$ survey:\\
1) Giant planets around giants are fairly common (about 10 $\%$). This is in
contrast to a frequency of $\approx$ 5 $\%$ for solar-type MS stars.\\
2) Planets around giant stars do not favour metal-rich stars (e.g. Pasquini
et al. 2007; Hekker $\&$ Melendez 2007; Hekker et al. 2008; Takeda et al. 
2008). A spectral analysis of the Tautenburg 
sample showed that the planet-hosting stars tend to be metal-poor 
(D\"ollinger 2008). This is in contrast to planet-hosting 
solar-type MS stars which tend to be metal-rich (e.g. Santos et al. 2004).\\
3) Planets around giant stars tend to be super planets with masses of
3--10 M$_{\mathrm{Jup}}$. For solar-type MS stars over half of the planets
have masses less than 3 M$_{\mathrm{Jup}}$. For giant stars (intermediate 
stellar mass) over half of the planets have masses more than 
3--5 M$_{\mathrm{Jup}}$.\\
4) Planets around giants have periods larger than $\sim$ 150~days.\\
5) Inner planets with orbital semimajor axes, $a$~$\leq$~0.7~AU are not
present (Johnson et al. 2007; Sato et al. 2008).

Giants are well suited to search for long-period planets around massive stars, 
but there are limitations on the possibility to detect short-period planets due to the fact that giants have large radii and they thus would have swallowed up 
any close-in planets. This region of the planetary orbital parameter space 
($P$~$<$~20~days) is thus inaccessible. 
In short, the results for planets around giant stars show different
characteristics to those found around solar-type (and presumably less massive)
MS stars.
 
\section{Data acquisition and analysis}

\begin{figure}[t]
\resizebox{\hsize}{!}{\includegraphics{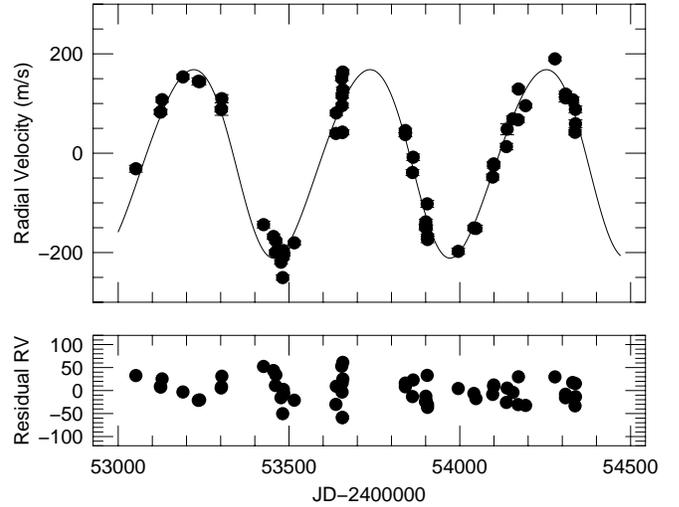}}
\caption{(Radial velocity measurements for 11~UMi. 
The solid line is the orbital solution (top). RV residuals
after subtracting the orbital solution (bottom).}
\end{figure}

Observations are described in D\"ollinger et al. (2007) and summarised here:
The data were acquired using the coud{\'e} {\'e}chelle spectrograph
of the 2m Alfred Jensch telescope, with a
resolving power of $R$ = 67{,}000. The wavelength
coverage was 4700--7400 {\AA} and the resulting signal-to-noise (S/N) ratio
typically greater than 150. 
Standard CCD data reduction
(bias-subtraction, flat-fielding and spectral extraction) was performed
using $IRAF$ routines. An iodine absorption cell placed
in the optical path provided the wavelength reference for the velocity
measurements.\\
The RVs were computed using standard procedures for measuring precise
stellar RVs using an iodine absorption cell (see Butler et al. 1996 and 
Endl et al. 2006).
Our spectral data was also used to derive important stellar parameters such 
as Fe abundance, surface gravity, and effective temperature.
For these a high S/N stellar spectrum taken without the 
iodine cell was used.
Stellar masses were derived using the online tool from Girardi
(http://stevoapd.inaf.it/cgi-bin/param) which is based on 
theoretical isochrones 
(Girardi et al. 2000) and a modified version of 
Jo$\!\!\!/$rgensen $\&$ Lindegren's (2005) method
(see da Silva et al. 2006 for a description). In a 
forthcoming paper (D\"ollinger 2009b) we will present in more detail
the chemical analysis and stellar parameter determination 
of all stars in our programme including 11~UMi and HD~32518.

\section{Results}
\subsection{11~UMi}
The stellar parameters of the K4~III star 
11~UMi (= HD~136726 = HR~5714 = HIP~74793) are 
summarised in Tab.~1. The stellar parameters like effective temperature,
T$_{\mathrm{eff}}$, Fe abundance, [Fe/H], logarithmic surface 
gravity, $\log g$, and microturbulent velocity, $\xi$ were 
derived from our spectral observations. Some of them like the stellar 
mass and radius comes from the Girardi isochrones. All other quantities were 
obtained from the $SIMBAD$ database.

\begin{table}[h]
\caption{Stellar parameters of 11~UMi.}
\vspace{-0.5cm}
$$
\begin{array}{lll}
\hline
\hline
\mathrm{Spectral\,\,type}     & \mathrm{K4III}        & $HIPPARCOS$\\
m_{\mathrm{V}}                & 5.024  \pm 0.005      & [\mathrm{mag}] \\
M_{\mathrm{V}}                & -0.37 \pm 0.12      & [\mathrm{mag}] \\
B-V                           & 1.395 \pm 0.005       & [\mathrm{mag}] \\
\mathrm{Parallax}             & 8.37 \pm 0.46         & [\mathrm{mas}] \\
\mathrm{Distance}             & 119.50 \pm 6.60        & [\mathrm{pc}] \\
M_{\mathrm{*}}^{(a)}          & 1.80 \pm 0.25       & [\mathrm{M_{\odot}}] \\
R_{\mathrm{*}}^{(a)}          & 24.08 \pm 1.84        & [\mathrm{R_{\odot}}] \\
\mathrm{Age}^{(a)}            & 1.56 \pm 0.54       & [\mathrm{Gyr}] \\
T_{\mathrm{eff}}^{(a)}        & 4340 \pm 70           & [\mathrm{K}] \\
\mathrm{[Fe/H]}^{(a)}         & +0.04 \pm 0.04       & [\mathrm{dex}] \\
\log{g}^{(a)}                 & 1.60 \pm 0.15          & [\mathrm{dex}] \\
\mathrm{micro\,turbulence}^{(a)}     & 1.6 \pm 0.8    & [\mathrm{{km\,s}^{-1}}]\\
\hline
\hline
\end{array}
$$
{\footnotesize
$^{(a)}$ D\"{o}llinger (2008), D\"{o}llinger (2009b), in preparation}
\end{table}

A total of 58 spectra with the iodine cell were obtained
for 11~UMi. These values are listed in Tab.~2.
The time series of the corresponding RV measurements is shown
in Fig.~1.

\begin{table}[h]
\caption{Radial velocity measurements for 11~UMi.}
\vspace{-0.5cm}
$$
\begin{array}{lrr}
\hline
\hline
\mathrm{JD}   & \mathrm{RV} [\mathrm{{m\,s}^{-1}}]  & \sigma [\mathrm{{m\,s}^{-1}}] \\
2\,453\,051.708713 &  -31.26  &   6.99  \\
2\,453\,123.534122 &  82.88   &  4.25  \\
2\,453\,124.544363 &  83.15   &  5.07  \\
2\,453\,128.572537 &  107.73  &   5.18  \\
2\,453\,189.537754 &  153.60  &   4.77  \\
2\,453\,234.499843 &  144.65  &   7.14  \\
2\,453\,238.357596 &  144.09  &   6.15  \\
2\,453\,301.421510 &  88.40   &  4.84  \\
2\,453\,302.506490 &  88.82   &  12.63  \\
2\,453\,303.341835 &  109.82  &   8.05  \\
2\,453\,425.520312 &  -143.80 &    6.56  \\
2\,453\,454.631345 &  -167.70 &    4.87  \\
2\,453\,460.388104 &  -199.74 &    5.17  \\
2\,453\,461.524772 &  -176.39 &    4.14  \\
2\,453\,476.501768 &  -219.40 &    3.85  \\
2\,453\,481.518826 &  -250.38 &    5.30  \\
2\,453\,482.527922 &  -209.20 &    4.60  \\
2\,453\,483.461198 &  -196.00 &    4.69  \\
2\,453\,484.469635 &  -203.42 &    3.77  \\
2\,453\,515.520600 &  -180.40 &    4.12  \\
2\,453\,637.321156 &  39.76   &  6.49  \\
2\,453\,638.391730 &  81.02   &  5.21  \\
2\,453\,654.389332 &  150.52  &   4.69  \\
2\,453\,655.379064 &  95.81   &  4.60  \\
2\,453\,655.596210 &  115.84  &   4.93  \\
2\,453\,656.401390 &  41.84   &  4.94  \\
2\,453\,656.405661 &  42.25   &  4.67  \\
2\,453\,657.375775 &  163.34  &   4.76  \\
2\,453\,657.599842 &  127.62  &   5.87  \\
2\,453\,899.479327 &  -148.17 &    5.69  \\
2\,453\,900.444086 &  -138.29 &    5.57  \\
2\,453\,901.507084 &  -150.42 &    5.15  \\
2\,453\,904.470771 &  -101.95 &    7.05  \\
2\,453\,905.494222 &  -173.77 &    6.25  \\
2\,453\,905.537161 &  -167.79 &    5.26  \\
2\,453\,840.454520 &  45.65   &   4.38  \\
2\,453\,840.532516 &  37.43   &  4.15  \\
2\,453\,861.463582 &  -38.92  &   6.01  \\
2\,453\,863.407248 &  -8.10   &  6.53  \\
2\,453\,995.331026 &  -197.52 &    6.12  \\
2\,454\,041.272281 &  -150.44 &    5.11  \\
2\,454\,047.246242 &  -151.41 &    6.31  \\
2\,454\,096.225428 &  -48.05  &   5.95  \\
2\,454\,136.425278 &  13.29   &  6.38  \\
2\,454\,138.554149 &  48.35   &  11.19  \\
2\,454\,155.590866 &  69.29   &  5.74  \\
2\,454\,099.228459 &  -24.55  &   5.54  \\
2\,454\,099.232683 &  -21.52  &   4.90  \\
2\,454\,192.593610 &  95.70   &  4.19  \\
2\,454\,170.642896 &  67.17   &  5.21  \\
2\,454\,171.669516 &  129.17  &   4.88  \\
2\,454\,278.472906 &  189.91  &   4.00  \\
2\,454\,309.397450 &  119.08 &   5.57  \\
2\,454\,309.401744 &  111.43 &   8.14  \\
2\,454\,330.340189 &  107.54 &   5.71  \\
2\,454\,337.321286 &  42.06  &  5.43  \\
2\,454\,338.344381 &  88.30  &  6.59  \\
2\,454\,338.545262 &  59.28  &  7.98 \\
\hline
\hline
\end{array}
$$
\end{table}

\begin{figure}
\resizebox{\hsize}{!}{\includegraphics{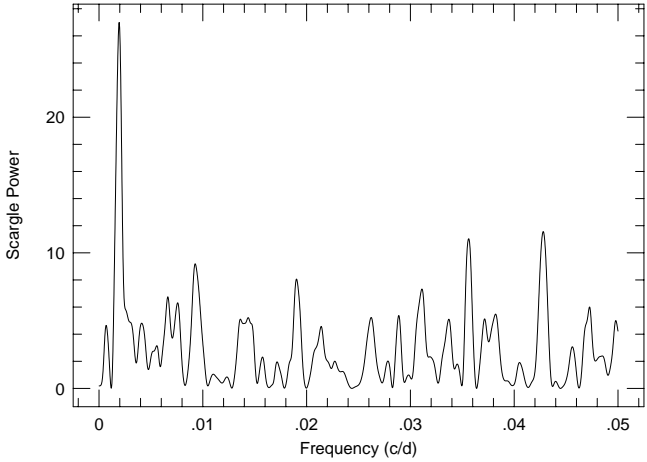}}
\caption{Scargle periodogram for 11~UMi. The high peak at Scargle power 
$\approx$ 23 is at a frequency $\nu$ = 0.00194 c\,d$^{-1}$ 
corresponding to a period of 518.18~days.
}
\end{figure}

\begin{figure}
\resizebox{\hsize}{!}{\includegraphics{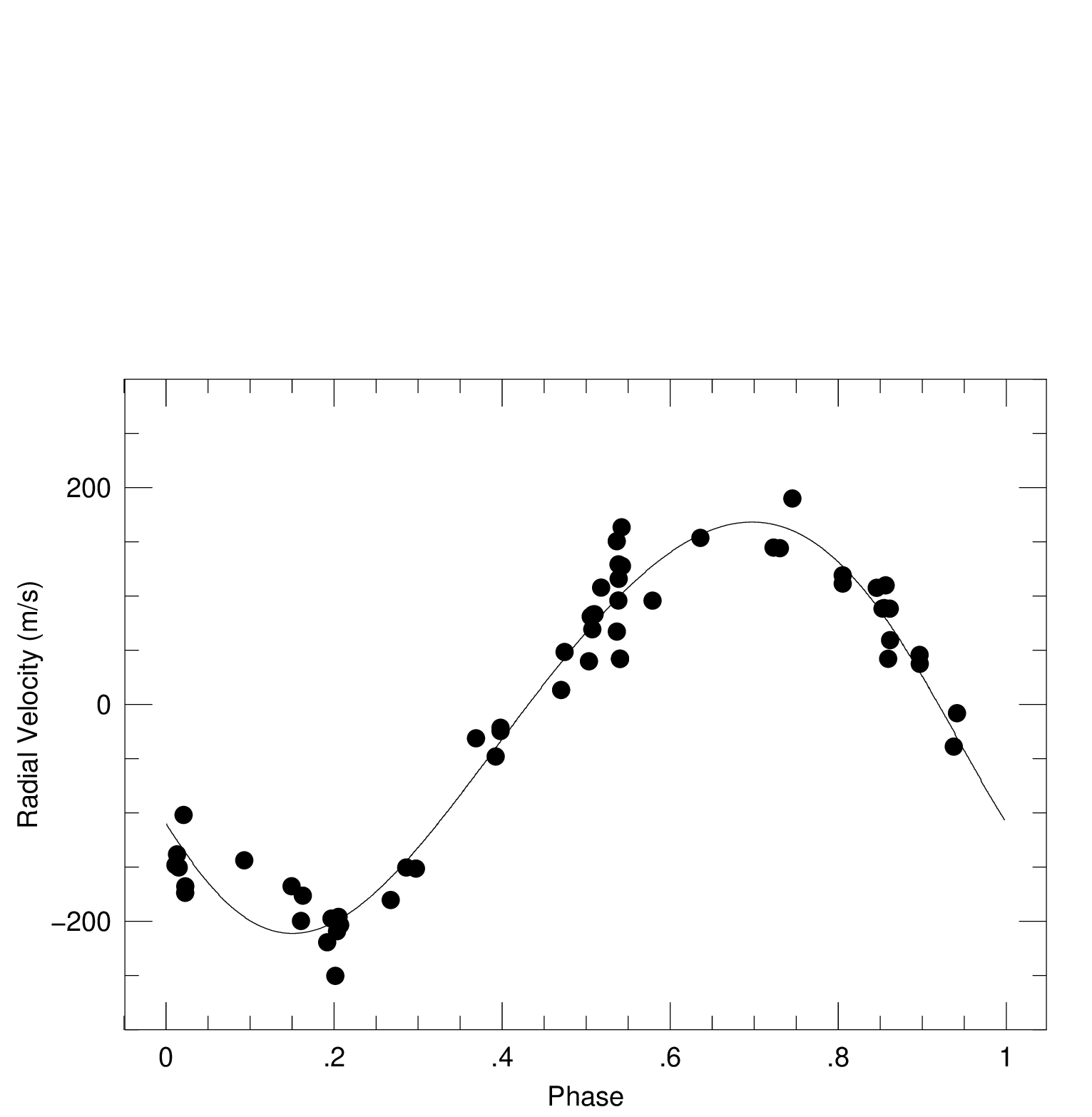}}
\caption{Radial velocity measurements for 11~UMi phased to the orbital period.
}
\end{figure}

A Scargle periodogram (Scargle 1982) was used to get an estimate of the
RV period that was used as an initial guess in the subsequent orbit fitting. 
Fig.~2 shows the Scargle periodogram of the 11~UMi RV measurements. There
is highly significant power ({\bf ``False Alarm Probability''}, FAP $\approx$
10$^{-10}$) at a frequency of $\nu$~=~0.00194~c\,d$^{-1}$ corresponding to 
a period of $P$~=~515~days. 

The parameters for the orbital solution to the RV data for 11 UMi are
listed in Tab.~3. The orbital fit to the data is shown
as a solid line in Fig.~1. The orbit is nearly circular.
The phase folded data and solution are shown in Fig.~3.
Using our derived stellar mass of 1.80 $\pm$ 0.245 M$_{\odot}$ results in a
minimum mass, $m \sin i$~=~11.20~$\pm$~2.47~M$_{\mathrm{Jup}}$.

\begin{figure}[h]
\resizebox{\hsize}{!}{\includegraphics{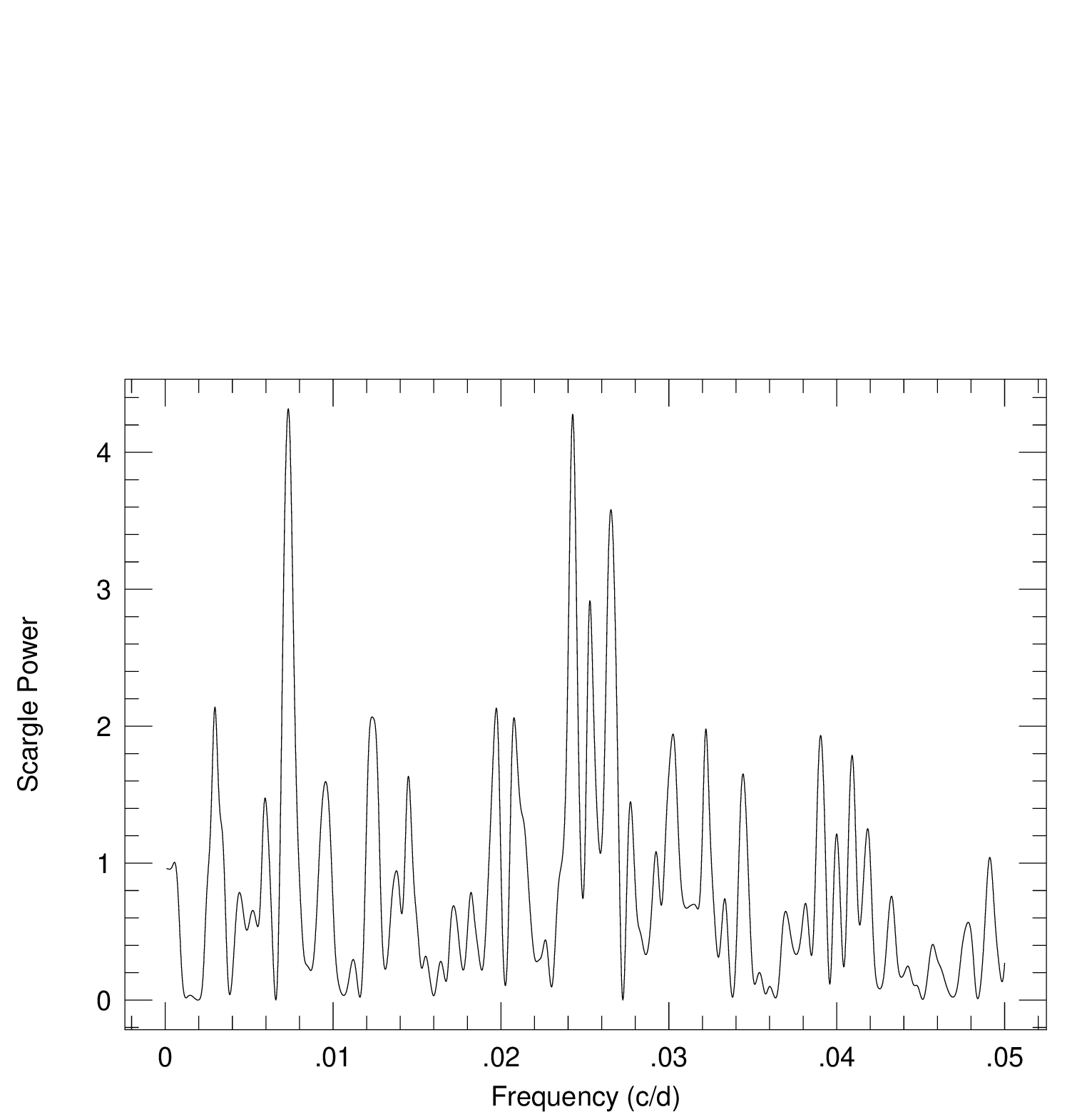}}
\caption{Scargle periodogram of the RV residuals of 11~UMi. There are no 
significant peaks in the residual RVs.
}
\end{figure}

\begin{table}[h]
\caption{Orbital parameters for the companion to 11~UMi.}
\vspace{-0.5cm}
$$
\begin{array}{lll}
\hline
\hline
\mathrm{Period} [\mathrm{days}]                       & 516.22 \pm 3.25\\
T_{\mathrm{periastron}}[\mathrm{JD}]                  & 52861.04 \pm 2.06\\
K [\mathrm{{m\,s}^{-1}}]                              & 189.70 \pm 7.15 \\
\sigma(\mathrm{O-C}) [\mathrm{{m\,s}^{-1}}]          & 27.50  \\
e                                                     & 0.08 \pm 0.03\\
\omega [\mathrm{deg}]                                 & 117.63 \pm 21.06\\
f(m) [\mathrm{M_{\odot}}]            & (3.61 \pm 0.44) \times 10^{-7} \\
a [\mathrm{AU}]                             & 1.54 \pm 0.07\\
\hline
\hline
\end{array}
$$
\end{table}

Fig.~4 shows the periodogram of the RV residuals 
(see lower panel Fig.1) after 
subtracting the orbital solution. There are no significant peaks present
out to a frequency of 0.05 c\,d$^{-1}$. A periodogram analysis
out to a higher frequency of 0.5 c\,d$^{-1}$ also reveals no significant
short-term variabilty that might be due to oscillations. This result is not
surprising since our sparse data sampling is inadequate for detecting such
short-period variability.

The rms scatter of the RV measurements about the orbital solution is
28 m\,s$^{-1}$, or a factor of 5 greater than the estimated
measurement error. This scatter most likely arises from stellar oscillations.
We can use the scaling relations of Kjeldsen $\&$ Bedding (1995) for p-mode
oscillations to estimate the velocity amplitude of such stellar oscillations.
Their Eq.~7 and the stellar parameters listed in Tab.~1
results in a RV amplitude of 27 m\,s$^{-1}$, comparable to our observed
scatter.

As with all giant stars we must be cautious about attributing any RV
variability to planetary companions. Such observed variability can also
arise from stellar surface structure as well. However, spots should produce 
variablity in other quantities. 

To test whether rotational modulation could account for the observed 
RV variability, we examined the $HIPPARCOS$ photometry. Fig.~5 shows the 
periodogram of the $HIPPARCOS$ photometry after removing outliers and taking
daily averages. There is no significant power at the observed orbital 
frequency. Fig.~6 shows the $HIPPARCOS$ photometry phase folded to the orbital 
period. There is no significant variability to a level of about 0.01 mag.

As an additional test we looked for variations in the
H$\alpha$, which can be an indicator of stellar activity. We measured
the line strength using a band pass of $\pm$ 0.6 {\AA} centered
on the core of the line, and two additional ones at $\pm$ 50 {\AA}
that provided a measurement of the continuum level (see D\"{o}llinger 2008)
for a more detailed description of how the H$\alpha$ was measured.

Fig.~7 shows the Scargle periodogram of the H$\alpha$ variations. There is no 
significant peak at the orbital frequency. In addition, Fig.~8 shows the 
H$\alpha$ indices as a function of time.  
However, there is also no significant feature in Fig.~8.

The rotational period estimated by the projected
rotational velocity, $v \sin i$~$=$~1.5~km\,s$^{-1}$ published by de Medeiros
and Mayor (1999), and the stellar radius, R$_{\mathrm{*}}$ listed in Tab.~1, is
P$_{\mathrm{rot}}$ $\leq$ 2$\pi$R$_{\mathrm{*}}$/($v \sin i$)~$\sim$~813$\pm$61~days,
which is incompatible with the observed period of the planetary companion
with a value of 516.22~days.

The lack of variability of photometric and H$\alpha$ data along with the 
exclusion of rotational modulation due to the incompatibility of orbital and 
rotational period strongly suggests
that the RV variations are due to Keplerian motion of a companion.

\begin{figure}[h]
\resizebox{\hsize}{!}{\includegraphics{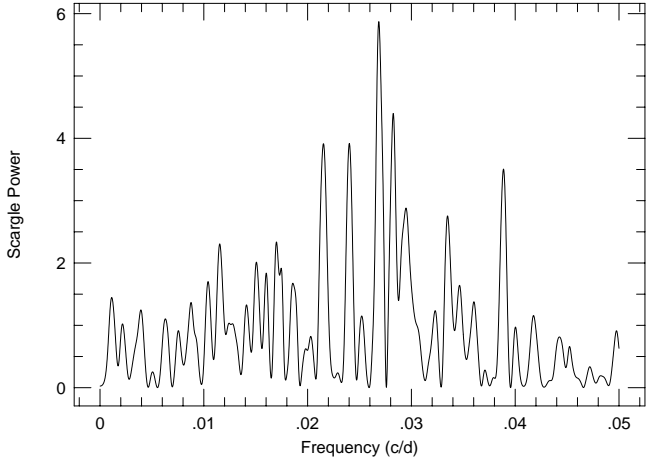}}
\caption{Scargle periodogram of the $HIPPARCOS$ photometry for 11~UMi.
}
\end{figure}

\begin{figure}[h]
\resizebox{\hsize}{!}{\includegraphics{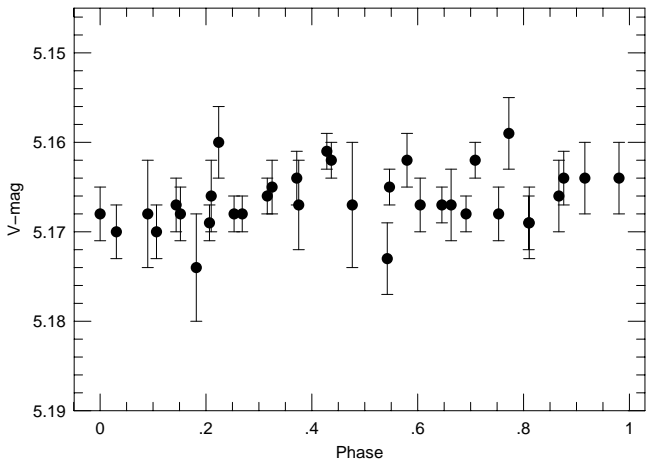}}
\caption{$HIPPARCOS$ photometry phased to the orbital period for 11 UMi.
}
\end{figure}

\begin{figure}[h]
\resizebox{\hsize}{!}{\includegraphics{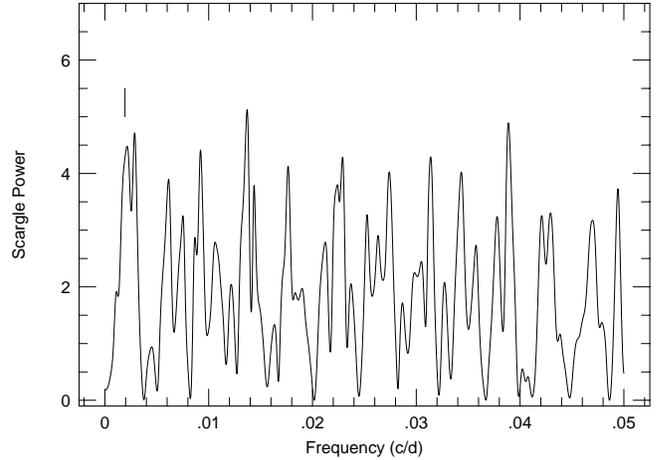}}
\caption{Scargle periodogram of the 11 UMi H$\alpha$ variations. There are
no significant frequencies in the data.
}
\end{figure}

\begin{figure}[h]
\resizebox{\hsize}{!}{\includegraphics{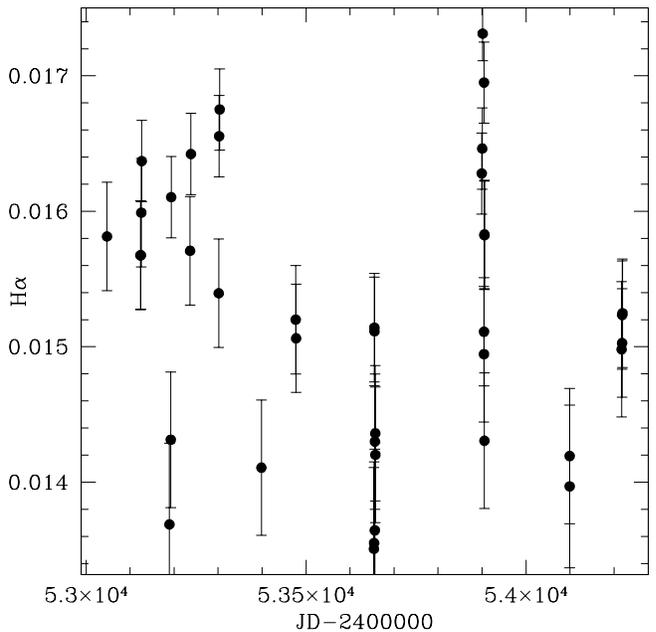}}
\caption{H$\alpha$ indices of 11 UMi as a function of time which show no significant feature.
}
\end{figure}

\subsection{HD 32518}
The stellar parameters for the K1~III giant HD~32518 (=  HR 1636 = HIP 24003) 
are listed in Tab.~4.

A total of 58 observations for this star were made with the iodine cell. These
are listed in Tab.~5. The top panel of Fig.~9 shows
the time series of the RV measurements. The Scargle periodogram of the
RV measurements is shown in Fig.~10. There is a significant power 
with FAP $\approx$ 10$^{-9}$ at a frequency of $\nu$~=~0.00634~c\,d$^{-1}$
corresponding to a period $P$~=~157~days.

\begin{table}[h]
\caption{Stellar parameters of HD 32518.}
\vspace{-0.5cm}
$$
\begin{array}{lll}
\hline
\hline
\mathrm{Spectral\,\,type}       & \mathrm{K1III}      & $HIPPARCOS$\\
m_{\mathrm{V}}                  & 6.436 \pm 0.005     & [\mathrm{mag}] \\
M_{\mathrm{V}}                  & 1.08 \pm 0.20     & [\mathrm{mag}] \\
B-V                             & 1.107 \pm 0.005     & [\mathrm{mag}] \\
\mathrm{Parallax}               & 8.52 \pm 0.78       & [\mathrm{mas}] \\
\mathrm{Distance}               & 117.40 \pm 10.70      & [\mathrm{pc}] \\
M_{\mathrm{*}}^{(a)}            & 1.13 \pm 0.18     & [\mathrm{M_{\odot}}] \\
R_{\mathrm{*}}^{(a)}            & 10.22 \pm 0.87      & [\mathrm{R_{\odot}}]] \\
\mathrm{Age}^{(a)}              & 5.83 \pm 2.58     & [\mathrm{Gyr}] \\
T_{\mathrm{eff}}^{(a)}          & 4580 \pm 70         & [\mathrm{K}] \\
\mathrm{[Fe/H]}^{(a)}           & -0.15 \pm 0.04     & [\mathrm{dex}] \\
\log{g}^{(a)}                   & 2.10 \pm 0.15        & [\mathrm{dex}] \\
\mathrm{micro\,turbulence}^{(a)} & 1.2 \pm 0.8        & [\mathrm{{km\,s}^{-1}}] \\
\hline
\hline
\end{array}
$$
{\footnotesize
$^{(a)}$ D\"{o}llinger (2008), D\"{o}llinger et al. (2009b), in preparation}
\end{table}

\begin{table}[h]
\caption{Radial velocity measurements for HD 32518.}
\vspace{-0.5cm}
$$
\begin{array}{lrr}
\hline
\hline
\mathrm{JD}   & \mathrm{RV} [\mathrm{{m\,s}^{-1}}]  & \sigma [\mathrm{{m\,s}^{-1
}}] \\
2\,453\,123.351010 &  105.08  &  7.43 \\
2\,453\,124.386724 &  85.34   & 11.09 \\
2\,453\,125.403781 &  72.24   & 9.95 \\
2\,453\,126.408572 &  91.46   & 12.92 \\
2\,453\,189.389311 &  -86.71  &  9.32 \\
2\,453\,193.436509 &  -140.78 &   15.46 \\
2\,453\,234.416771 &  -110.55 &   11.18 \\
2\,453\,236.346655 &  -65.43&    14.54 \\
2\,453\,238.613668 &  -79.66&    15.01 \\
2\,453\,251.378491 &  14.36 &   7.73 \\
2\,453\,301.517243 &  99.85 &   5.88 \\
2\,453\,302.451461 &  88.70 & 7.43 \\
2\,453\,303.592656 &  120.72  &14.98 \\
2\,453\,390.500274 &  -120.54 &   11.75 \\
2\,453\,419.656517 &  45.57 &   13.40 \\
2\,453\,420.328859 &  40.07 &   8.11 \\
2\,453\,460.365367 &  64.99 &   9.53 \\
2\,453\,461.365880 &  66.52 &   10.70 \\
2\,453\,477.306512 &  18.02 &   7.13 \\
2\,453\,477.318085 &  12.94 &   8.68 \\
2\,453\,481.394615 &  6.01  &7.29 \\
2\,453\,482.334865 &  -8.85  &  6.55 \\
2\,453\,483.334070 &  -9.94  &  5.83 \\
2\,453\,484.342026 &  -9.12  &  7.54 \\
2\,453\,633.523740 &  34.30  &  7.01 \\
2\,453\,633.527953 &  26.92  &  7.02 \\
2\,453\,599.345222 &  92.59  &  8.01 \\
2\,453\,627.435618 &  48.95  &  7.24 \\
2\,453\,750.466481 &  85.62  &  9.73 \\
2\,453\,751.528063 &  72.02  &  7.74 \\
2\,453\,786.272842 &  59.49  &6.48 \\
2\,453\,654.296462 &  -77.45 &   8.84 \\
2\,453\,654.609583 &  -70.65 &   7.78 \\
2\,453\,655.281384 &  -54.55 &   9.20 \\
2\,453\,655.565684 &  -67.12 &   7.02 \\
2\,453\,656.286804 &  -49.87 &   10.50 \\
2\,453\,656.581996 &  -85.58 &   5.80 \\
2\,453\,657.280835 &  -87.84 &   7.14 \\
2\,453\,657.591340 &  -86.29 &   7.85 \\
2\,453\,899.376857 &  43.58  &  9.75 \\
2\,453\,900.350466 &  97.07  &  10.77 \\
2\,453\,901.372687 &  50.58  &  17.22 \\
2\,453\,904.363223 &  112.97 &   16.96 \\
2\,453\,904.408987 &  79.03  &13.74 \\
2\,453\,905.358215 &  114.71  &  13.37 \\
2\,453\,905.528678 &  61.39   & 10.39 \\
2\,454\,041.684294 &  13.09   & 9.80 \\
2\,454\,041.696783 &  11.51   & 6.77 \\
2\,454\,052.404390 &  61.57   & 7.79 \\
2\,454\,018.480256 &  -53.35  &  6.50 \\
2\,454\,134.599349 &  -74.57  &  11.37 \\
2\,454\,134.607821 &  -85.48  &  7.21 \\
2\,454\,136.403558 &  -96.73  &  8.35 \\
2\,454\,155.560424 &  -92.16  &  8.29 \\
2\,454\,186.280654 &  -24.96  &  6.88 \\
2\,454\,248.370147 &  93.64   & 8.71 \\
2\,454\,337.568079 &  -114.32 &   8.47 \\
2\,454\,338.486736 &  -97.49  &  11.97 \\
\hline
\hline
\end{array}
$$
\end{table}

\begin{figure}[h]
\resizebox{\hsize}{!}{\includegraphics{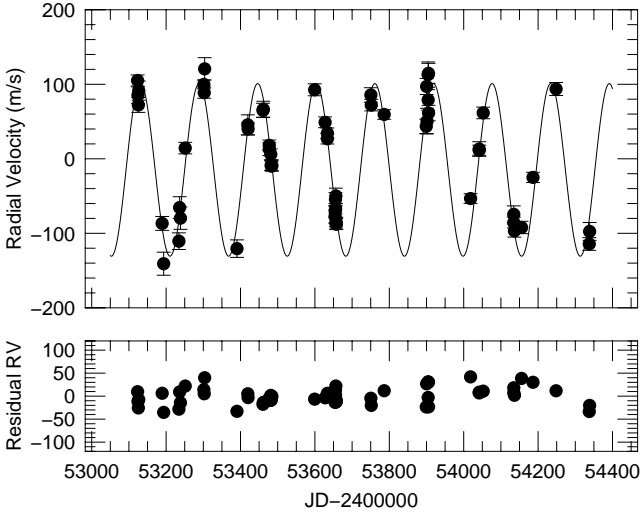}}
\caption{Radial velocity measurements for HD 32518. The solid line represents
the orbital solution (top). RV residuals after subtracting the orbital solution (bottom).
}
\end{figure}

\begin{figure}[h]
\resizebox{\hsize}{!}{\includegraphics{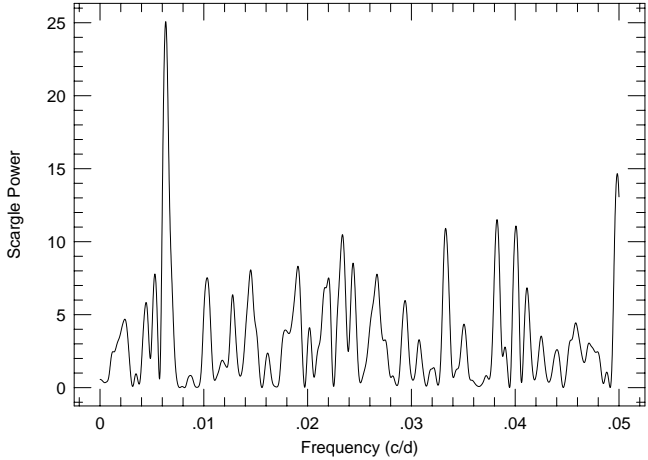}}
\caption{Scargle periodogram for HD 32518. The high peak at a frequency of 
$\nu$ = 0.00634 c\,d$^{-1}$ corresponds to a period of 157.73 days.}
\end{figure}

\begin{figure}[h]
\resizebox{\hsize}{!}{\includegraphics{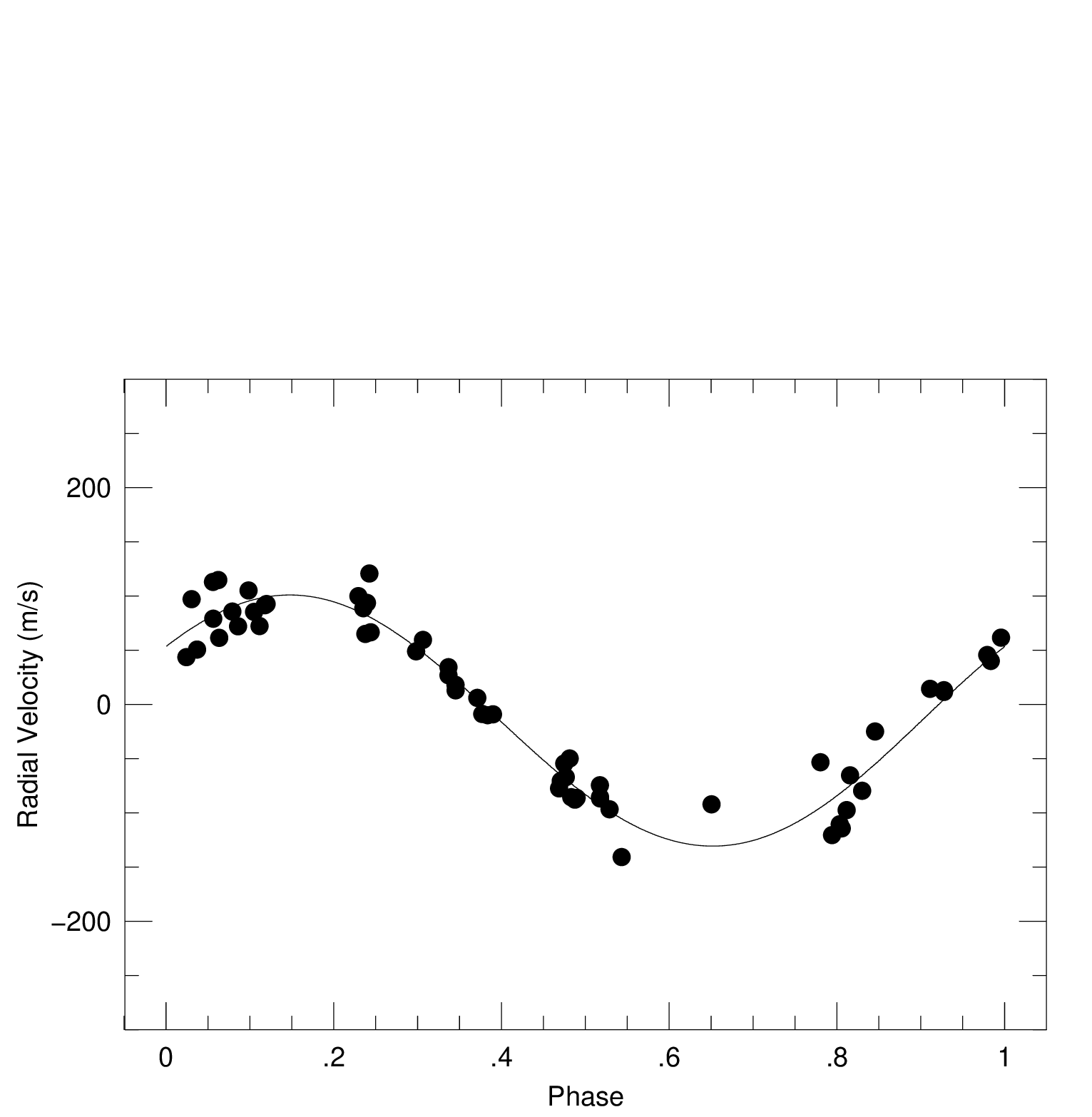}}
\caption{Radial velocity measurements for HD 32518 phased to the orbital 
period. The line represents the orbital solution.}
\end{figure}

An orbital solution to the RV data yields a period, 
$P$~=~157.54~$\pm$~0.38~days and a circular orbit, $e$~=~0.008~$\pm$~0.032.
The orbital solution is shown as a line in Fig.~9.
All of the orbital parameters are listed in Tab.~6.
Fig.~11 shows the RV variations phase-folded to the orbital period.

Our stellar mass estimate of 1.13$\pm$0.18~M$_{\odot}$ for HD 32518 results 
in a minimum companion mass of 3.04~$\pm$~0.68~M$_{\mathrm{Jup}}$.

\begin{figure}[h]
\resizebox{\hsize}{!}{\includegraphics{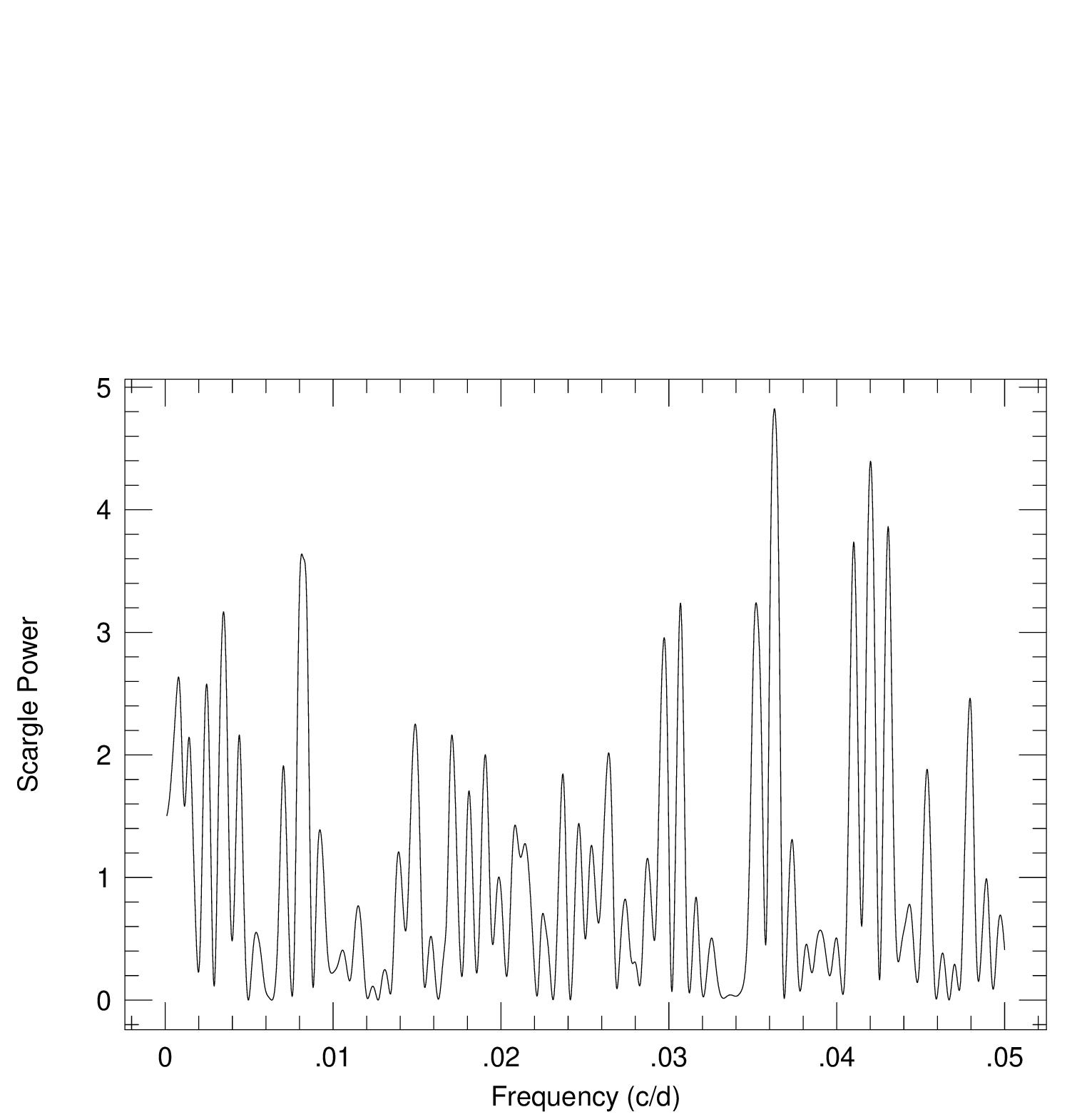}}
\caption{Scargle periodogram of the RV residuals of HD 32518. There is no significant peak in the RV residuals.}
\end{figure}

Once again, the rms scatter of about 18~m\,s$^{-1}$ can largely be 
explained by stellar oscillations. Using the Kjeldsen $\&$ Bedding (1995)
scaling relations results in a RV amplitude of about 27~m\,s$^{-1}$ for
the predicted stellar oscillations in HD 32518. 

\begin{table}[h]
\caption{Orbital parameters for the companion to HD 32518.}
\vspace{-0.5cm}
$$
\begin{array}{lll}
\hline
\hline
\mathrm{Period} [\mathrm{days}]                     & 157.54 \pm 0.38 \\
T_{\mathrm{periastron}}[\mathrm{JD}]                & 52950.29 \pm 13.66\\
K [\mathrm{{m\,s}^{-1}}]                            & 115.83 \pm 4.67 \\
\sigma(\mathrm{O-C)}) [\mathrm{{m\,s}^{-1}}]        & 18.33 \\
e                                                   & 0.01 \pm 0.03\\
\omega [\mathrm{deg}]                               & 306.11 \pm 126.71\\
f(m) [\mathrm{M_{\odot}}]  & (2.20 \pm 0.24) \times 10^{-8} \\
a [\mathrm{AU}]                                     & 0.59 \pm 0.03\\
\hline
\hline
\end{array}
$$
\end{table}

Fig.~12 shows the periodogram of the RV residuals (lower panel of Fig.~9) after 
subtracting the orbital solution. There are no significant 
peaks present out to a frequency of 0.05 c\,d$^{-1}$. An extension of the 
periodogram out to higher frequencies (0.5 c\,d$^{-1}$) also reveals no 
additional periodic RV variations. 

To determine the nature of the RV variations, we again examined the 
$HIPPARCOS$ photometry (see Fig.~13) and our H$\alpha$ measurements.
Fig.~14 shows the Scargle periodogram of the daily averages of the 
$HIPPARCOS$ photometry with outliers removed. There are no
significant peaks near the orbital frequency marked by the vertical line.

\begin{figure}[h]
\resizebox{\hsize}{!}{\includegraphics{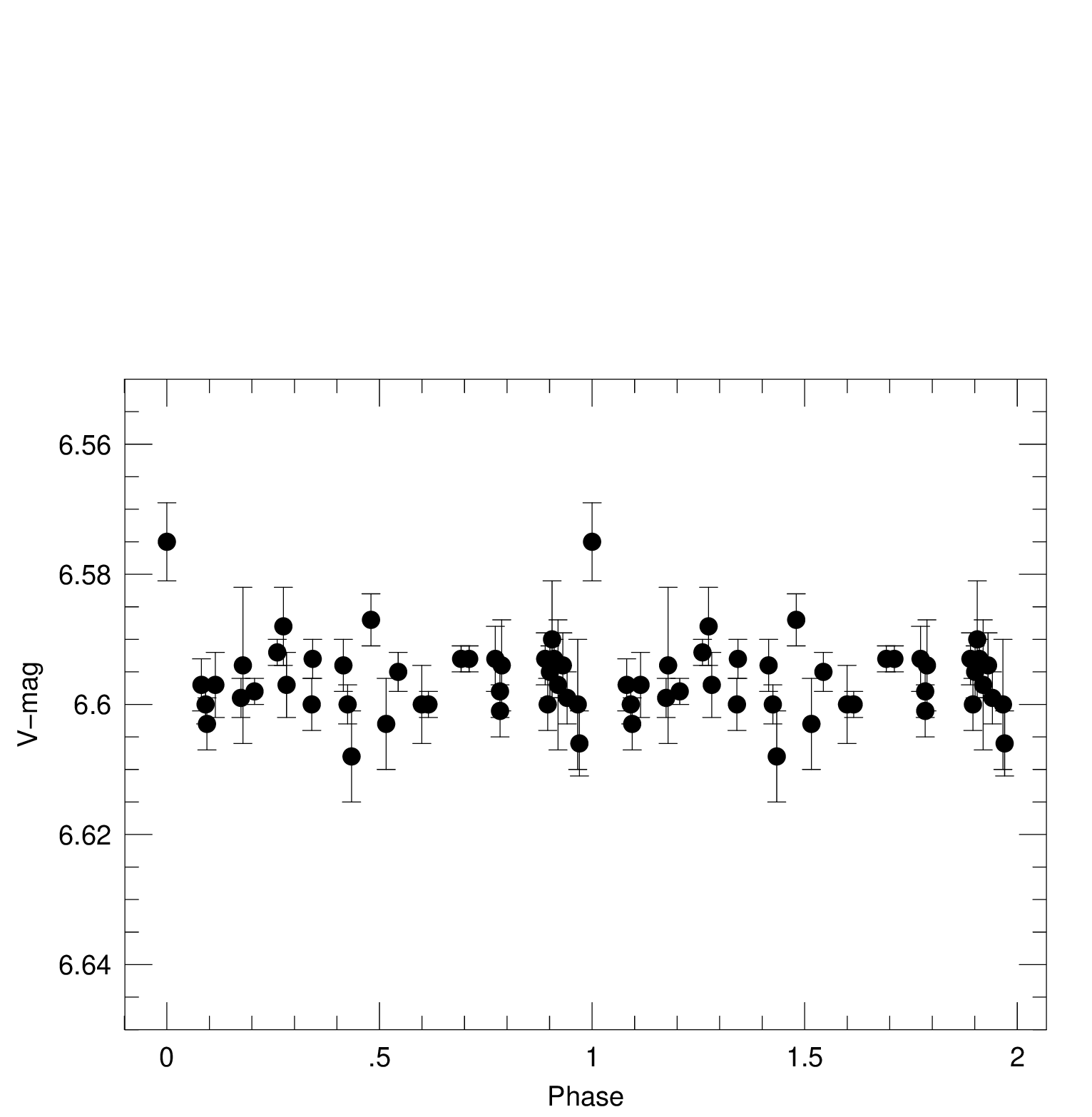}}
\caption{$HIPPARCOS$ photometry for HD 32518 phased to the orbital period.}
\end{figure}

\begin{figure}[h]
\resizebox{\hsize}{!}{\includegraphics{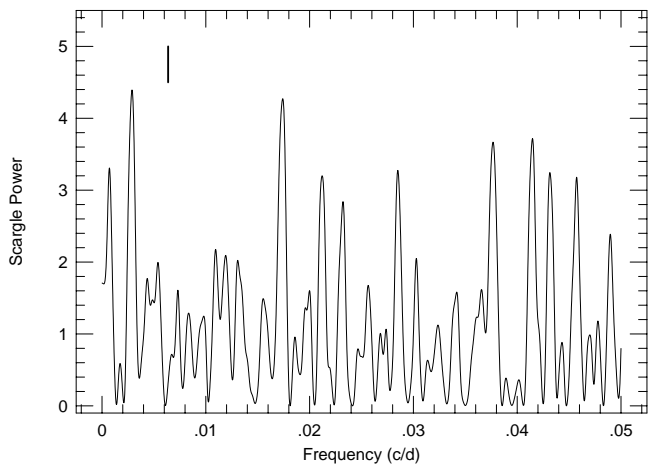}}
\caption{Scargle periodogram of the $HIPPARCOS$ photometry for HD 32518.
The vertical line marks the orbital frequency. The $HIPPARCOS$ photometry shows
no significant peak at this frequency.}
\end{figure}

\begin{figure}[h]
\resizebox{\hsize}{!}{\includegraphics{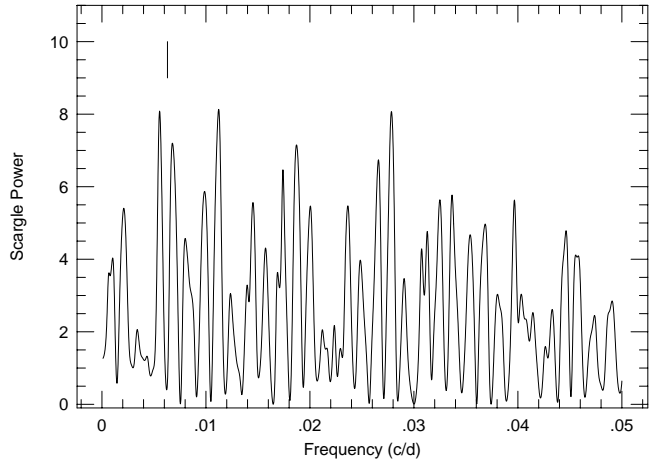}}
\caption{Scargle periodogram of HD~32518 H$\alpha$ variations. The vertical line marks again the orbital frequency. The H$\alpha$ variations show no significant peak at the orbital frequency of the planetary companion.}
\end{figure}

\begin{figure}[h]
\resizebox{\hsize}{!}{\includegraphics{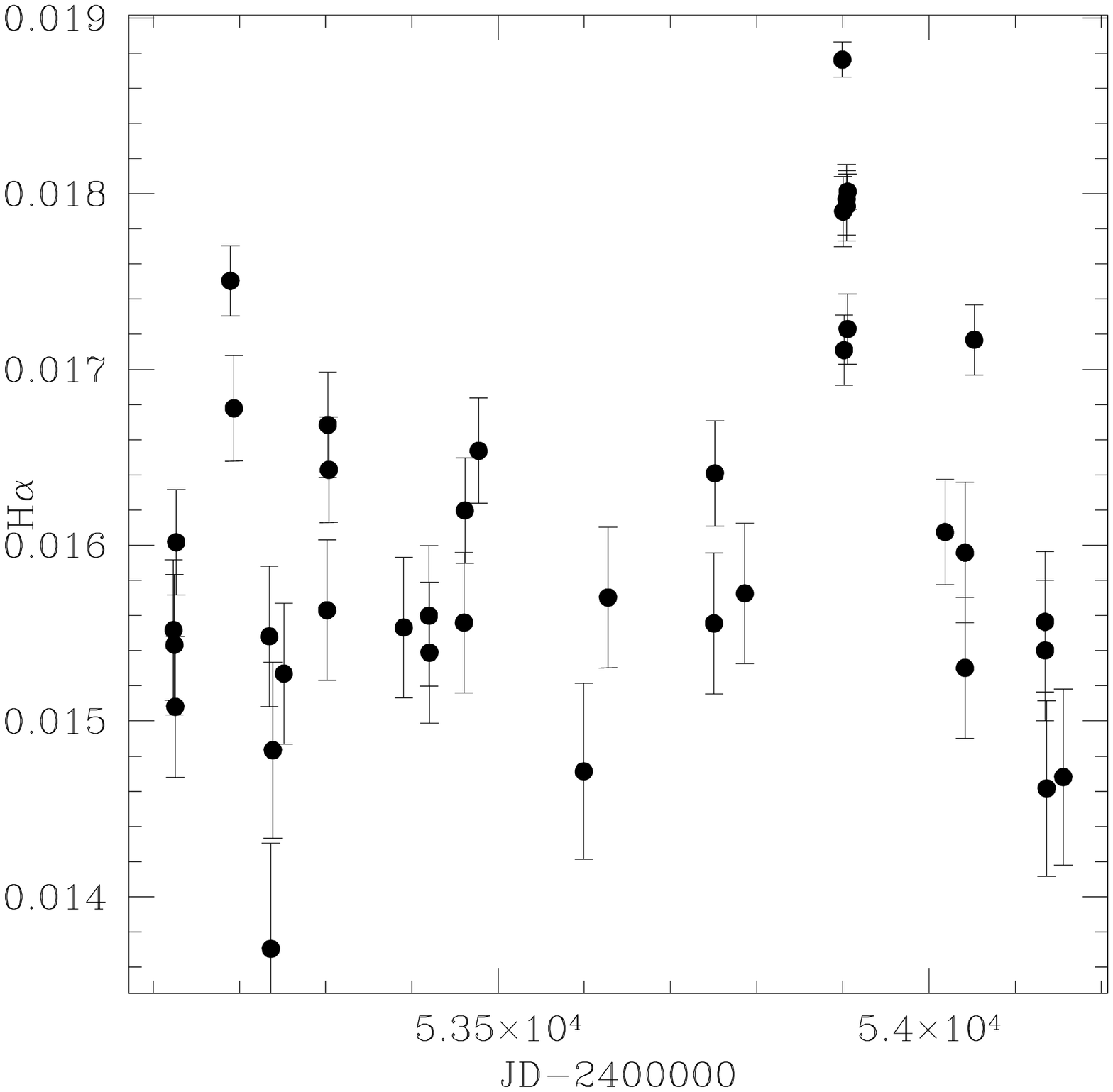}}
\caption{H$\alpha$ indices of HD~32518 as a function of time.}
\end{figure}

The Scargle periodogram of the H$\alpha$ variability is
shown in Fig.~15. There is no significant peak at the orbital
frequency. Fig.~16 shows the H$\alpha$ indices of HD~32518 as a function of 
time.

From the projected rotational velocity, $v \sin i$~$=$~1.2~km\,s$^{-1}$
published by de Medeiros and Mayor (1999), and the adopted stellar radius
(see Tab. 4) we have estimated the upper limit of the rotational period.
This calculated value of 430$\pm$40~days is different from the orbital period
(see Tab. 6).
 
The lack of photometric and H$\alpha$ variability with the RV period
and the different rotational period suggests that the RV variations for 
this star are due to a planetary companion.

\section{Discussion}
Since the discovery of 51 Peg b in 1995, we know about the presence of 
exoplanets around solar-type stars. In the meantime the detected number of 
extrasolar planets has increased tremendously. At the moment more than 350 
have been discovered mostly using the RV method which is currently the 
method of choice for planet-hunting.\\
In the past solar-type MS stars were favoured targets for
planet searches, and consequently most of the published planets orbit
this type of host star. Very famous in this context are the so-called
``hot Jupiters'', Jupiter-like planets which are in very close orbits
around their parent stars. Their existence was a big surprise and is still a
puzzle. In contrast to MS stars this kind of exoplanet will normally not 
be present around evolved stars with their enlarged envelope because the 
planetary companions would be swallowed up. Therefore planets around giants 
have periods larger than $\sim$ 150 days with exception of $\xi$ Aql 
(Sato et al. 2008) which shows an orbital period of 136.75 days.\\
The planet of HD 32518 with a period of 157.24 days is slightly above this 
limit. HD 32518 b and 11 UMi b have a nearly circular orbit. In this case
the variations in the RV curves can be mimiced by surface structures
like starspots. However, the lack of variability in the $HIPPARCOS$ 
and the H$\alpha$ data for both giants is more consistent with the planet 
hypothesis. We caution the reader that the $HIPPARCOS$ photometry was
not simultaneous with our RV measurements. Thus, we cannot exclude that spots
were not present when $HIPPARCOS$ observed these stars, but are present now 
and are causing the RV variations. However, our H$\alpha$ measurements were 
made simultaneously to the RV data. We therefore believe that
the detected periods of several hundred days in both stars are not due to
rotational modulation, but rather to planetary companions.\\
During the $TLS$ programme we have found at least 6 planetary companion 
candidates. This corresponds to an occurence rate of around 10 $\%$ for 
giant stars, which is in opposite to around 5 $\%$ for MS stars. 
More recently 3 additional objects are found, which would
bring the percentage to 15 $\%$. This higher frequency of planet occurrence
around evolved stars seems to be consistent with recent theoretical 
predictions.
Kennedy $\&$ Kenyon (2008) used semi-analytical disk models to show that
the probability of a star having at least one giant planet rises
linearly from 0.4 to 3 M$_{\odot}$. They predict that the frequency
of giant planets is about 10 $\%$ for 1.5 M$_{\odot}$ stars, consistent with 
our initial estimate.\\
11 UMi has a nearly solar metal abundance, [Fe/H] = $+0.04$ $\pm$ 0.04 dex 
while HD 32518 is slightly metal-poor with a value of $-0.15$ $\pm$ 0.04 dex. 
However, both stars are relatively 
``metal-poor'' compared to previous results of planet-hosting MS stars which 
tend to be metal-rich (Santos et al. 2004), but of higher abundance compared 
to other planet-hosting giant stars which tend to be metal-poor 
(Schuler et al. 2005; Pasquini et al. 2007).\\  
Our stellar mass determinations indicate that the MS
progenitor to HD 32518 was most likely a late F-type star. More interesting
is 11 UMi whose stellar mass suggests a progenitor that was an early
A-type star. Intriguingly, the more massive star of the two has the more
massive substellar companion ($m \sin i$~=~10.5~M$_{\mathrm{Jup}}$ compared
to 3.04~M$_{\mathrm{Jup}}$. This is consistent with the observed trend that 
more massive stars tend to have more massive planets, but more statistics
are needed to confirm this. Comparing the results of other searches for
planets around giant stars Johnson et al. (2007) as well as Lovis and Mayor 
(2007) also found that more massive stars seem to harbour more massive 
planetary systems (see their Fig.~11).
A possible explanation for this behaviour can maybe delivered by model 
predictions (Laughlin et al. 2004; Ida $\&$ Lin 2005). According to them giant 
planet formation depends on the 
mass and surface density of the protoplanetary disc besides the metallicity.
For these parameters the mass of the star plays a key role in the sense
that more massive stars will have more massive disks and higher surface
densities, which enables to accrete larger amounts of material.      

\begin{acknowledgements}
We are grateful to the user support group of the Alfred Jensch telescope:
B. Fuhrmann, J. Haupt, Chr. H\"{o}gner, U. Laux, M. Pluto, J. Schiller,
and J. Winkler.
This research made use of the $SIMBAD$ database, operated at CDS, Strasbourg,
France.
\end{acknowledgements}

\end{document}